\documentclass[11pt,a4paper]{article}
\usepackage{amssymb}

\def\onehalf{\textstyle{\frac{1}{2}}}

\def\D{{\mathcal D}{}}
\def\Gammabol{{\stackrel{\circ}{\Gamma}}{}}

\def\Abol{{\stackrel{~\circ}{A}}{}}

\def\Rbol{{\stackrel{\circ}{R}}{}}
\def\Lbol{{\stackrel{\circ}{\mathcal L}}{}}

\def\Dbol{{\stackrel{\circ}{\mathcal D}}{}}

\def\Gammaw{{\stackrel{\bullet}{\Gamma}}{}}

\def\Aw{{\stackrel{~\bullet}{A}}{}}

\def\Rw{{\stackrel{\bullet}{R}}{}}

\def\jw{{\stackrel{\bullet}{\jmath}}{}}
\def\iw{{\stackrel{\bullet}{\imath}}{}}
\def\tw{{\stackrel{\bullet}{t}}{}}
\def\L{{\mathcal L}{}}
\def\Lw{{\stackrel{\bullet}{\mathcal L}}{}}
\def\Tw{{\stackrel{\bullet}{T}}{}}

\def\Kw{{\stackrel{\bullet}{K}}{}}

\def\Dw{{\stackrel{\bullet}{\mathcal D}}{}}

\def\sw{{\stackrel{\bullet}{S}}{}}

\def\be{\begin{equation}}
\def\ee{\end{equation}}
\def\ba{\begin{eqnarray}}
\def\ea{\end{eqnarray}}

\begin{document}

\renewcommand{\thefootnote}{\arabic{footnote}}
\noindent
{\Large \bf Consistent Gravitationally-Coupled Spin-2 Field Theory}
\vskip 0.7cm
\noindent
{\bf H. I. Arcos$^a$, Tiago Gribl Lucas$^b$ and J. G. Pereira$^b$}
\vskip 0.3cm \noindent
$^a${\it Universidad Tecnol\'ogica de Pereira\\
\null \hskip 0.2cm A.A. 97, La Julita, Pereira, Colombia}
\vskip 0.2cm \noindent
$^b${\it Instituto de F\'{\i}sica Te\'orica, UNESP-Univ Estadual Paulista \\
\null \hskip 0.2cm Caixa Postal 70532-2, 01156-970 S\~ao Paulo, Brazil}

\vskip 0.8cm
\begin{quote}
{\bf Abstract.~}{\footnotesize Inspired by the translational gauge structure of teleparallel gravity, the theory for a fundamental massless spin-2 field is constructed. Accordingly, instead of being represented by a symmetric second-rank tensor, the fundamental spin-2 field is assumed to be represented by a spacetime (world) vector field assuming values in the Lie algebra of the translation group. The flat-space theory naturally emerges in the Fierz formalism and is found to be equivalent to the usual metric-based theory. However, the gravitationally coupled theory, with gravitation itself described by teleparallel gravity, is shown not to present the consistency problems of the spin-2 theory constructed on the basis of general relativity.

}
\end{quote}

\vskip 0.4cm \noindent

\section{Introduction}
\label{intro}

It is well known that higher ($s>1$) spin fields, and in particular a spin-2 field, present consistency problems when coupled to gravitation \cite{deser1, deser2}. The problem is that the divergence identities satisfied by the field equations of a spin-2 field in Minkowski spacetime are no longer valid when it is coupled to gravitation. In addition, the coupled equations are no longer gauge invariant. The basic underlying difficulty is related to the fact that the covariant derivative of general relativity---which defines the gravitational coupling prescription---is non-commutative, and this introduces unphysical constraints on the spacetime curvature.

Due to the fact that linearized gravity represents a spin-2 field, the dynamics of a fundamental spin-2 field in Minkowski space, according to the usual approach \cite{deser3}, is expected to coincide with the dynamics of a linear perturbation of the metric around flat spacetime:\footnote{We are going to use the Greek alphabet $\mu, \nu, \rho, \dots = 0,1,2,3$ to denote indices related to spacetime, also known as world indices. The first half of the Latin alphabet $a,b,c, \dots = 0,1,2,3$ will be used to denote algebraic indices related to the tangent spaces, each one a Minkowski spacetime with metric $\eta_{ab} = \mathrm{diag}(+1,-1,-1,-1)$.}
\be
g_{\mu \nu} = \eta_{\mu \nu} + \psi_{\mu \nu}.
\ee
For this reason, a fundamental spin-2 field is usually assumed to be described by a rank-two, symmetric tensor $\psi_{\mu \nu} = \psi_{\nu \mu}$. However, conceptually speaking, this is not the most fundamental notion of a spin-2 field. As is well known, although the gravitational interaction of scalar and vector fields can be described in the metric formalism, the gravitational interaction of spinor fields requires a tetrad formalism \cite{dirac}. The tetrad formalism can then be considered to be more fundamental than the metric formulation in the sense that it is able to describe the gravitational interaction of both tensor and spinor fields. Accordingly, the tetrad field can be said to be more fundamental than the metric.

Relying on this property, instead of similar to a linear perturbation of the metric, a fundamental spin-2 field should be considered a linear perturbation of the tetrad field. Tetrads can be better understood in the context of the tangent bundle. It is formed by spacetime (the base space), and by all Minkowski tangent spaces (the fibers of the bundle) that exist attached to each point of spacetime. Denoting a general tetrad field by $h^a{}_\mu$, the spacetime and the tangent space metrics are related by
\be
g_{\mu \nu} = h^a{}_\mu h^b{}_\nu \, \eta_{ab}.
\label{getahh}
\ee
Due to the existence of this relation between the fiber and the spacetime metrics, the tangent bundle is said to be soldered \cite{KoNu}. In absence of gravitation the tetrad is trivial (closed 1-form), and $g_{\mu \nu}$ represents the Minkowski metric in a general coordinate system.

Denoting by $e^a{}_\mu$ a trivial tetrad representing the Minkowski metric, a fundamental spin-2 field $\phi^a{}_\mu$ should therefore be defined by
\be
h^a{}_\mu = e^a{}_\mu + \phi^a{}_\mu.
\label{hPertu}
\ee
Since the tetrad is a translational-valued vector field, $\phi^a{}_\mu$ will also be a translational-valued vector field,
\be
\phi_\mu = \phi^a{}_\mu \, P_a,
\ee
with $P_a = \partial_a$ the translation generators. Observe that, in the usual {\em metric} formulation of gravity, the symmetry of the metric tensor eliminates six degrees of freedom of the sixteen original degrees of freedom of $g_{\mu \nu}$. In the {\em tetrad} formulation, on the other hand, local Lorentz invariance is responsible for eliminating six degrees of freedom of the sixteen original degrees of freedom of $h^a{}_\mu$, yielding the same number of independent components of $g_{\mu \nu}$. Of course, the same equivalence must hold in relation to the fields $\psi_{\mu \nu}$ and $\phi^a{}_\mu$.

Now, as is well known, in teleparallel gravity the gravitational field is represented by a translational gauge potential $B_\mu = B^a{}_\mu P_a$, which appears as the nontrivial part of the tetrad field \cite{review}. This means essentially that $\phi^a{}_\mu$ is similar to the gauge potential of teleparallel gravity. Accordingly, its dynamics must coincide with the dynamics of linearized teleparallel gravity. On account of these considerations, the purpose of this paper is to construct a teleparallel-based field theory for a fundamental spin-2 field. This will be done according to the following scheme. In section~\ref{cofra}, for the sake of completeness, we review the basic notions of connections and frames. In section \ref{TG}, we give a brief description of teleparallel gravity, also known as teleparallel equivalent of general relativity. Then, by using this theory as paradigm, we construct in section \ref{FlatTheory} the field theory for a fundamental spin-2 field in flat spacetime. Owing to the (abelian) gauge structure of teleparallel gravity, it naturally emerges in the Fierz formulation, being quite similar to electromagnetism, a gauge theory for the $U(1)$ group. In section \ref{s2plusGra}, by using the gravitational coupling prescription appropriate for a translational-valued vector field, we obtain the corresponding spin-2 theory in the presence of gravitation. In addition to present duality symmetry, it is gauge and local Lorentz invariant. This means that, even in the presence of gravitation, $\phi^a{}_\mu$ has the correct number of independent components to represent a massless spin-2 field. In section \ref{s2Source}, we consider the spin-2 field as source of gravitation, and we obtain the covariant conservation law of the true source of gravitation. Due to the fact that the spin connection of teleparallel gravity represents inertial effects only---not gravitation---the covariant derivative appearing in this conservation law is commutative. As a consequence, no constraints on the background geometry show up, yielding a fully consistent gravitationally-coupled spin-2 field theory. Finally, in section \ref{FR}, we sum up the results obtained.

\section{Lorentz connections and frames}
\label{cofra}

A basic ingredient of any relativistic theory is the spin connection $A_\mu$, a connection assuming values in the Lie algebra of the Lorentz group,
\be
A_\mu = \onehalf \, A^{ab}{}_\mu \, S_{ab},
\ee
with $S_{ab} = - S_{ba}$ a given representation of the Lorentz generators. The corresponding Lorentz covariant derivative is given by the Fock-Ivanenko operator $\D_\mu$ \cite{fi}, which acting on a Lorentz vector field $\phi^a$, for example, assumes the form
\be
\D_\mu \phi^a = \partial_\mu \phi^a + A^{a}{}_{b \mu} \, \phi^b.
\ee
Analogously to relation (\ref{getahh}), to each spin connection $A^{a}{}_{b \mu}$ there corresponds a linear connection $\Gamma^{\rho}{}_{\nu \mu}$ given by \cite{livro}
\be
\Gamma^{\rho}{}_{\nu \mu} = h_{a}{}^{\rho} \partial_{\mu} h^{a}{}_{\nu} +
h_{a}{}^{\rho} A^{a}{}_{b \mu} h^{b}{}_{\nu} \equiv h_{a}{}^{\rho} \D_{\mu} h^{a}{}_{\nu}.
\label{geco}
\ee
The inverse relation is
\be
A^{a}{}_{b \mu} =
h^{a}{}_{\nu} \partial_{\mu}  h_{b}{}^{\nu} +
h^{a}{}_{\nu} \Gamma^{\nu}{}_{\rho \mu} h_{b}{}^{\rho} \equiv h^{a}{}_{\nu} \nabla_{\mu} h_{b}{}^{\nu},
\label{gsc}
\ee
with $\nabla_{\mu}$ the covariant derivative in the connection $\Gamma^{\nu}{}_{\rho \mu}$. Relations (\ref{geco}) and (\ref{gsc}) are different ways of expressing the property that the total covariant derivative---that is, with connection terms for both indices---of the tetrad vanishes identically:
\be
\partial_{\mu} h^{a}{}_{\nu} - \Gamma^{\rho}{}_{\nu \mu} h^{a}{}_{\rho} +
A^{a}{}_{b \mu} h^{b}{}_{\nu} = 0.
\label{todete}
\ee
Since the last index of the spin connection is a tensorial index, one can write
\be
A^{a}{}_{b c} = A^{a}{}_{b \mu} \, h_c{}^\mu.
\label{Aabc}
\ee 

The curvature and torsion of $A^{a}{}_{b \mu}$ are defined respectively by
\be
R^a{}_{b \nu \mu} = \partial_{\nu} A^{a}{}_{b \mu} -
\partial_{\mu} A^{a}{}_{b \nu} + A^a{}_{e \nu} A^e{}_{b \mu}
- A^a{}_{e \mu} A^e{}_{b \nu}
\ee
and
\be
T^a{}_{\nu \mu} = \partial_{\nu} h^{a}{}_{\mu} -
\partial_{\mu} h^{a}{}_{\nu} + A^a{}_{e \nu} h^e{}_{\mu}
- A^a{}_{e \mu} h^e{}_{\nu}.
\label{tordef}
\ee
Using relations (\ref{geco}) and (\ref{gsc}), they can be expressed in a purely spacetime form as
\be
\label{sixbm}
R^\rho{}_{\lambda\nu\mu} = \partial_\nu \Gamma^\rho{}_{\lambda \mu} -
\partial_\mu \Gamma^\rho{}_{\lambda \nu} +
\Gamma^\rho{}_{\eta \nu} \Gamma^\eta{}_{\lambda \mu} -
\Gamma^\rho{}_{\eta \mu} \Gamma^\eta{}_{\lambda \nu}
\ee
and
\be
T^\rho{}_{\nu \mu} =
\Gamma^\rho{}_{\mu\nu}-\Gamma^\rho{}_{\nu\mu}.
\label{sixam}
\ee

A general Cartan connection $\Gamma^\rho{}_{\mu\nu}$ can always be decomposed in the form \cite{KoNu}
\be
\Gamma^\rho{}_{\mu\nu} = {\stackrel{\circ}{\Gamma}}{}^{\rho}{}_{\mu \nu} +
K^\rho{}_{\mu\nu},
\label{prela0}
\ee
where\footnote{All quantities related to general relativity will be denoted with an over ``$\circ$''.}
\be
{\stackrel{\circ}{\Gamma}}{}^{\sigma}{}_{\mu \nu} = {\textstyle
\frac{1}{2}} g^{\sigma \rho} \left( \partial_{\mu} g_{\rho \nu} +
\partial_{\nu} g_{\rho \mu} - \partial_{\rho} g_{\mu \nu} \right)
\label{lci}
\ee
is the Levi-Civita connection, and
\be
K^\rho{}_{\mu\nu} = {\textstyle
\frac{1}{2}} \left(T_\nu{}^\rho{}_\mu + T_\mu{}^\rho{}_\nu -
T^\rho{}_{\mu\nu}\right) 
\label{contor}
\ee
is the contortion tensor. Using relation (\ref{geco}), the decomposition (\ref{prela0}) can be rewritten as
\be
A^a{}_{b\nu} = \Abol^a{}_{b\nu} + K^a{}_{b\nu},
\label{rela00}
\ee
where $\Abol^a{}_{b \nu}$ is the spin connection of general relativity.

The quantities
\be
h^a = h^a{}_\mu \, dx^\mu \quad \mbox{and} \quad
h_a = h_a{}^\mu \, \partial_\mu
\ee
represent Lorentz frames. These frames are classified according to the value of the coefficient of anholonomy $f^{c}{}_{a b}$, which is de\-fined by the commutation relation
\begin{equation}
[h_{a}, h_{b}] = f^{c}{}_{a b}\ h_{c}.
\label{eq:comtable}
\end{equation}
As a simple calculation shows, it is given by
\be
f^a{}_{cd} = h_c{}^\mu \, h_d{}^\nu (\partial_\nu
h^a{}_\mu - \partial_\mu h^a{}_\nu).
\ee
For example, in special relativity, the class of inertial frames is defined by all frames for which $f^{c}{}_{a b} = 0$. They are called, for this reason, holonomic frames. The anholonomy of these frames is entirely related to the inertial forces present in those frames. Starting with an inertial frame, different classes of frames are obtained by performing {\em local} (point dependent) Lorentz transformations. Inside each class, different frames are related through {\em global} (point independent) Lorentz transformations. Similarly, in the presence of gravitation, there is also a preferred class of frames: the class whose anholonomy is related to gravitation only, not with inertial effects. This class of frames, therefore, reduces to the inertial class when gravitation is switched off.

\section{Fundamentals of teleparallel gravity}
\label{TG}

Teleparallel gravity corresponds to a gauge theory for the translation group. Its main property is to be completely equivalent to general relativity. Although equivalent, however, the two theories are conceptually different. In general relativity, curvature is used to {\it geometrize} the gravitational interaction. In this theory, geometry replaces the concept of gravitational force, and the trajectories are determined, not by force equations, but by geodesics. In teleparallel gravity, on the other hand, gravitation is attributed to torsion, but in this case torsion accounts for gravitation not by geometrizing the interaction, but by acting as a {\it force}. In consequence, there are no geodesics in teleparallel gravity, but only force equations quite analogous to the Lorentz force equation of electrodynamics \cite{LorFor}. This difference is consistent with the gauge structure of teleparallel gravity.
 
One may wonder why gravitation has two equivalent descriptions. This duplicity is related to universality of free fall. Gravitation has, like the other fundamental interactions of nature, a description in terms of a gauge theory---just teleparallel gravity. Universality of free fall, however, makes it possible a geometrized description, given by general relativity. As the sole universal interaction of nature, it is the only one to allow a geometrical interpretation, and consequently two alternative descriptions. One may also wonder why a gauge theory for the translation group, and not for other spacetime group. The reason is related to the source of gravitation, that is, energy and momentum. As is well known from Noether's theorem, these quantities are conserved provided the source Lagrangian is invariant under spacetime translations. It is then natural to expect that the gravitational field be associated to the translation group. This is similar to the electromagnetic field, whose source---the electric four-current---is conserved due to the invariance of the source Lagrangian under transformations of the unitary group $U(1)$, the gauge group of Maxwell's theory.

A crucial point of teleparallel gravity is that, in the class of frames $h'_b$ which re\-duces to the inertial class in absence of gravitation, its spin connection\footnote{All quantities related to teleparallel gravity will be denoted with an over ``$\bullet$''.} vanishes eve\-ry\-where \cite{review}: 
\be
\Aw'^{a}{}_{b \mu} = 0.
\label{wsc}
\ee
In a Lorentz rotated frame $h^a = \Lambda^a{}_b(x) \, h'^b$, remembering that a general spin connection $A^{a}{}_{b \mu}$ transforms according to
\be
A^{a}{}_{b \mu} = \Lambda^{a}{}_{c}(x) \, A'^{c}{}_{d \mu} \,
\Lambda_{b}{}^{d}(x) + \Lambda^{a}{}_{c}(x) \, \partial_{\mu} \Lambda_{b}{}^{c}(x),
\label{ltsc}
\ee
it assumes the form
\be
\Aw^{a}{}_{b \mu} = \Lambda^{a}{}_{e}(x) \, \partial_\mu \Lambda_{b}{}^{e}(x).
\label{ltwsc}
\ee
We see clearly from this expression that the teleparallel spin connection represents purely inertial effects, not gravitation \cite{gemt3}. In fact, as an easy calculation shows, its curvature vanishes identically:
\be
\Rw^a{}_{b \mu \nu} = 0.
\ee

In the class of frames $h'_b$, the tetrad of teleparallel gravity is written as
\be
h'^a{}_\mu = \partial_\mu x'^a + B'^a{}_\mu,
\ee
where $B'^a{}_\mu$ is the translational gauge potential. In the general frame $h^a$, it assumes the form
\be
h^a{}_\mu = \Dw_\mu x^a + B^a{}_\mu,
\label{TetraB}
\ee
where $B^a{}_\mu = \Lambda^a{}_b \, B'^b{}_\mu$, and
\be
\Dw_\mu x^a = \partial_\mu x^a + \Aw^a{}_{b \mu} \, x^b.
\ee
For a tetrad with a non-trivial gauge potential $B^a{}_\mu$, torsion is non-vanishing:
\be
\Tw^a{}_{\mu \nu} =
\Dw_\mu h^a{}_\nu - \Dw_\nu h^a{}_\mu \neq 0.
\label{Torsion1}
\ee
This equation can be rewritten in the form
\be
\Tw^a{}_{\mu \nu} = h^a{}_\lambda \left(\Gammaw^\lambda{}_{\nu \mu} -
\Gammaw^\lambda{}_{\mu \nu} \right).
\ee
where
\be
\Gammaw^\lambda{}_{\mu \nu} = h_a{}^\lambda \, \Dw_\nu h^a{}_\mu
\ee
is the so-called Weitzenb\"ock connection. 
In contrast to general relativity, therefore, in which gravitation is represented by curvature, in teleparallel gravity it is represented by torsion. 

Denoting  $h = \det (h^a{}_\mu)$ and $k = 8 \pi G/c^4$, the gravitational Lagrangian of teleparallel gravity is written as \cite{maluf}
\be
\Lw = \frac{h}{4 k} \; \sw_{a}{}^{\rho \sigma} \; \Tw^a{}_{\rho \sigma},
\label{tela}
\ee
where
\be
\sw_a{}^{\rho \sigma} \equiv - \,
\frac{k}{h} \, \frac{\partial {\Lw}}{\partial (\partial_\sigma h^a{}_{\rho})} = h_a{}^\nu \left( \Kw^{\rho \sigma}{}_{\nu} - \delta_\nu{}^{\sigma} \;  \Tw^{\theta \rho}{}_{\theta} + \delta_\nu{}^{\rho} \; \Tw^{\theta \sigma}{}_{\theta} \right)
\label{supote}
\ee
is the superpotential, with $\Kw^{\rho \sigma}{}_{\nu}$ the teleparallel contortion. Using relation (\ref{prela0}) for the specific case of teleparallel torsion, it is possible to verify that
\begin{equation}
\Lw = \Lbol - \partial_\mu \left(2 \, h \, k^{-1} \,
\Tw^{\nu \mu}{}_\nu \right),
\end{equation}
where
\begin{equation}
\Lbol = - \frac{\sqrt{-g}}{2 k} \; \Rbol,
\end{equation}
is the Einstein-Hilbert Lagrangian of general relativity. Up to a divergence, therefore, the teleparallel Lagrangian is equivalent to the Einstein-Hilbert Lagrangian of general relativity. 

Variation of the Lagrangian (\ref{tela}) yields the sourceless gravitational field equation, which in teleparallel gravity appears naturally in the potential form \cite{Moller},
\be
\partial_\sigma (h \sw_a{}^{\rho \sigma}) -
k \, h \, \jw_{a}{}^{\rho} = 0,
\label{tfe0}
\ee
where
\be
\jw_{a}{}^{\rho} \equiv - \, \frac{1}{h} \frac{\partial {\Lw}}{\partial h^a{}_{\rho}} =
\frac{1}{k} \, h_a{}^{\lambda} \, \sw_c{}^{\nu \rho} \,
\Tw^c{}_{\nu \lambda} - \frac{h_a{}^{\rho}}{h} \, \Lw +
\frac{1}{k} \, \Aw^c{}_{a \sigma} \sw_c{}^{\rho \sigma}
\label{ptem10bis}
\ee
is the gravitational energy-momentum pseudo-current. Through a lengthy, but straightforward calculation, it is possible to show that
\begin{equation}
\partial_\sigma (h \sw_a{}^{\rho \sigma}) -
k \, h \, \jw_{a}{}^{\rho} =
h \left({\stackrel{\circ}{R}}_a{}^{\rho} -
\onehalf \, h_a{}^{\rho} \;
{\stackrel{\circ}{R}} \right).
\end{equation}
As expected due to the equivalence between the corresponding Lagrangians, the teleparallel field equation is equivalent to Einstein equation. Owing to the anti-symmetry of the su\-per\-potential in the last two indices, the gravitational energy-momentum pseudo-current is conserved in the ordinary sense:
\be
\partial_\rho (h \jw_{a}{}^{\rho}) = 0.
\label{iplustCon}
\ee

Now, in teleparallel gravity, the gravitational energy-momentum pseudo-current can be decomposed according to
\be
\jw_{a}{}^{\rho} = \tw_{a}{}^{\rho} + \iw_{a}{}^{\rho},
\ee
where
\be
\tw_{a}{}^{\rho} =
\frac{1}{k} \, h_a{}^{\lambda} \, \sw_c{}^{\nu \rho} \,
\Tw^c{}_{\nu \lambda} - \frac{h_a{}^{\rho}}{h} \, \Lw
\label{graem}
\ee
is a {tensorial current that represents the energy-momentum of gravity alone} \cite{gemt}, and
\be
\iw_{a}{}^{\rho} = \frac{1}{k} \, \Aw^c{}_{a \sigma} \sw_c{}^{\rho \sigma}
\label{InerEM}
\ee
is the energy-momentum current of inertia \cite{gemt3}. Of course, by its very nature, the latter is non-covariant. Using the fact that
\be
\partial_\sigma (h \sw_a{}^{\rho \sigma}) -
\Aw ^c{}_{a \sigma} (h \, \sw_{c}{}^{\rho \sigma}) \equiv \Dw_\sigma (h \sw_a{}^{\rho \sigma}),
\label{DotFI}
\ee
the field equation (\ref{tfe0}) can be rewritten in the form
\be
\Dw_\sigma (h \sw_a{}^{\rho \sigma}) -
k \, h \, \tw_{a}{}^{\rho} = 0.
\label{fe11}
\ee

Then comes the crucial point: since the teleparallel spin con\-nec\-tion (\ref{ltwsc}) has vanishing curvature, the corresponding Fock-Ivanenko derivative is com\-mutative:
\be
[\Dw_\rho , \Dw_\sigma] = 0.
\ee
Taking into account the anti-symmetry of the su\-per\-potential in the last two indices, it follows from the field equation (\ref{fe11}) that the tensorial current (\ref{graem}) is conserved in the  covariant sense:
\be
\Dw_\rho (h \tw_{a}{}^{\rho}) = 0.
\label{GravEMcon}
\ee
Of course, as it does not represent the total energy-momentum density---in the sense that the energy-momentum density of inertia is not included---it does not need to be truly conserved.

\section{Spin-2 field in flat spacetime}
\label{FlatTheory}

Using teleparallel gravity as paradigm, we proceed now to the construction of a theory for a fundamental spin-2 field in flat spacetime.

\subsection{Gauge transformations}

In the inertial frame $e'^a$, the tetrad describing the flat Minkowski spacetime is of the form
\be
e'^a{}_\mu = \partial_\mu x'^a.
\label{TriTetra}
\ee
A spin-2 field $\phi'^a{}_\mu$ corresponds to a linear perturbation of this tetrad:
\be
h'^a{}_\mu = \partial_\mu x'^a + \phi'^a{}_\mu.
\label{NonTriTetra}
\ee
In this class of frames, therefore, the vacuum is represented by
\be
\phi'^a{}_\mu = \partial_\mu \xi^a(x),
\ee
with $\xi^a(x)$ an arbitrary function of the spacetime coordinates $x^\rho$. In fact, such $\phi'^a{}_\mu$ represents simply a gauge transformation
\be
x'^a \to x'^a + \xi^a(x)
\ee
of the tangent space (or fiber) coordinates. This means that the gauge transformation associated to the spin-2 field $\phi^a{}_\mu$ is
\be
\phi'^a{}_\mu \to \phi'^a{}_\mu - \partial_\mu \xi^a(x).
\label{GaugeTrans}
\ee
Of course, $h'^a{}_\mu$ is invariant under such transformations.

\subsection{Field strength and Bianchi identity}

The next step towards the construction of a field theory for $\phi'^a{}_\mu$ is to define the analogous of torsion: 
\be
F'^a{}_{\mu \nu} = \partial_\mu \phi'^a{}_{\nu} - \partial_\nu \phi'^a{}_{\mu}.
\label{FS}
\ee
This tensor is actually the spin-2 field-strength. As can be easily verified, $F'^a{}_{\mu \nu}$ is gauge invariant. Furthermore, it satisfies the Bianchi identity
\be
\partial_\rho F'^a{}_{\mu \nu} + \partial_\nu F'^a{}_{\rho \mu} + \partial_\mu F'^a{}_{\nu \rho} = 0,
\label{BI}
\ee
which can equivalently be written in the form
\be
\partial_\rho (\varepsilon^{\lambda \rho \mu \nu} \, F'^a{}_{\mu \nu}) = 0,
\label{BI2}
\ee
with $\varepsilon^{\lambda \rho \mu \nu}$ the totally anti-symmetric, flat spacetime Levi-Civita tensor.

\subsection{Lagrangian and field equation}

Considering that the dynamics of a spin-2 field must coincide with the dynamics of linear gravity, its Lagrangian will be similar to the Lagrangian (\ref{tela}) of teleparallel gravity. One has just to replace the teleparallel torsion $\Tw^a{}_{\mu \nu}$ by the spin-2 field strength $\sqrt{k} \, F'^a{}_{\mu \nu}$. It is interesting to notice that when we do that, the spin-2 analogous of the teleparallel superpotential $\sw^a{}_{\mu \nu}$ is the Fierz tensor \cite{FP}
\be
{\mathcal F}'_a{}^{\mu \nu} = e'_a{}^\rho \, {\mathcal K}'^{\mu \nu}{}_\rho - e'_a{}^\nu \, e'_b{}^\rho \, F'^{b \mu}{}_\rho + e'_a{}^\mu \, e'_b{}^\rho \, F'^{b \nu}{}_\rho,
\label{SuperFree}
\ee
 with
\be
{\mathcal K}'^{\mu \nu}{}_\rho = \onehalf \left(e'_a{}^\nu \, F'^{a \mu}{}_\rho +
e'^a{}_\rho \, F'_a{}^{\mu \nu} - e'_a{}^\mu \, F'^{a \nu}{}_\rho \right)
\ee
the spin-2 analogous of the teleparallel contortion. Considering that in the absence of gravitation $\det(e'^a{}_\mu) = 1$, the Lagrangian for a massless spin-2 field is
\be
\L' = \frac{1}{4} \, {\mathcal F}'_a{}^{\mu \nu} \, F'^a{}_{\mu \nu}.
\label{EleLinha}
\ee
By performing variations with respect to $\phi^a{}_\rho$, we obtain
\be
\partial_\mu {\mathcal F}'_a{}^{\rho \mu} = 0.
\label{eom linear}
\ee
This is the field equation satisfied by a massless spin-2 field in Minkowski spacetime, as seen from the inertial frame $e'^a{}_\mu$. Notice that teleparallel gravity naturally yields the Fierz formulation for a spin-2 field \cite{linearizedtg}.

\subsection{Duality symmetry}

The spin-2 field can be viewed as an Abelian gauge field with the {\em internal} index replaced by an {\em external} Lorentz index. Due to the presence of the tetrad, Lorentz and spacetime indices can be transformed into each other. As a consequence, its Hodge dual will necessarily include additional index contractions in relation to the usual dual. Taking into account all possible contractions, its dual turns out to be given by \cite{geneDual}
\be
^\star F'^a{}_{\mu \nu} = \onehalf \epsilon_{\mu \nu \rho \sigma} \, {\mathcal F}'^{a \rho \sigma}.
\label{geneHd}
\ee
Substituting the Fierz tensor (\ref{SuperFree}), we find
\be
^\star F'^a{}_{\mu \nu} = \onehalf \epsilon_{\mu \nu \rho \sigma} \left( F'^{a \rho \sigma} -
e'^{a \sigma} e'_b{}^\lambda \, F'^{b \rho}{}_\lambda +
e'^{a \rho} e'_b{}^\lambda \, F'^{b \sigma}{}_\lambda \right).
\label{geneHd2}
\ee

Let us now consider the Bianchi identity (\ref{BI2}). Written for the dual of $F^a{}_{\mu \nu}$, it reads
\be
\partial_\rho (\varepsilon^{\lambda \rho \mu \nu} \, ^\star F'^a{}_{\mu \nu}) = 0.
\label{BI3}
\ee
Substituting $^\star F'^a{}_{\mu \nu}$ as given by Eq.~(\ref{geneHd}), we get
\be
\partial_\rho {\mathcal F}'^{a \mu \rho} = 0,
\ee
which is the field equation (\ref{eom linear}). We see in this way that, provided the generalized Hodge dual (\ref{geneHd}) for soldered bundles is used, the spin-2 field has duality symmetry. This is actually an expected result because the dynamics of a spin-2 field must coincide with the dynamics of linear gravity, which has already been shown to present duality symmetry \cite{linearizedgr}. We remark finally that, in the usual Fierz formulation of a spin-2 field, identity (\ref{BI2}) has to be put by hand in order to get the correct number of equations \cite{novello}. In the present formulation it appears as a consequence of considering the spin-2 field as a perturbation of the tetrad instead of the metric.

\subsection{Passage to a general frame}

In a Lorentz rotated frame $e^a = \Lambda^a{}_b(x) \, e'^a$, the tetrad assumes the form
\be
e^a{}_\mu \equiv \Dw_\mu x^a = \partial_\mu x^a + \Aw^a{}_{b \mu} \, x^b.
\ee
In this frame, the vacuum of $\phi^a{}_\mu$ turns out to be represented by
\be
\phi^a{}_\mu = \Dw_\mu \xi^a(x),
\ee
whereas the gauge transformations assume the form
\be
\phi^a{}_\mu \to \phi^a{}_\mu - \Dw_\mu \xi^a(x).
\label{GenGT}
\ee
The field strength (\ref{FS}), on the other hand, becomes
\be
F^a{}_{\mu \nu} = \Dw_\mu \phi^a{}_{\nu} - \Dw_\nu \phi^a{}_{\mu}.
\label{FS2}
\ee
Accordingly, the Bianchi identity reads
\be
\Dw_\rho F^a{}_{\mu \nu} + \Dw_\nu F^a{}_{\rho \mu} + \Dw_\mu F^a{}_{\nu \rho} = 0,
\label{BI3bis}
\ee
which is equivalent to
\be
\Dw_\rho(\varepsilon^{\lambda \rho \mu \nu} \, F^a{}_{\mu \nu}) = 0.
\ee
Analogously, the Fierz tensor turns out to be
\be
{\mathcal F}_a{}^{\mu \nu} = e_a{}^\rho \, {\mathcal K}^{\mu \nu}{}_\rho - e_a{}^\nu \, e_b{}^\rho \, F^{b \mu}{}_\rho + e_a{}^\mu \, e_b{}^\rho \, F^{b \nu}{}_\rho.
\ee
Now, as a simple inspection shows, the Lagrangian (\ref{EleLinha}) is invariant under local Lorentz transformations, that is,
\be
\L' \equiv \L = \frac{1}{4} \, {\mathcal F}_a{}^{\mu \nu} \, F^a{}_{\mu \nu}.
\ee
The corresponding field equation,
\be
\Dw_\mu {\mathcal F}_a{}^{\rho \mu} \equiv \partial_\mu {\mathcal F}_a{}^{\rho \mu} -
\Aw^b{}_{a \mu} \, {\mathcal F}_b{}^{\rho \mu} = 0,
\label{eom linear2}
\ee
represents the field equation satisfied by a massless spin-2 field in Minkowski spacetime, as seen from the general frame $e^a{}_\mu$. It is clearly invariant under the gauge transformation (\ref{GenGT}). It is important to observe that the theory has twenty two constraints: sixteen of the invariance under the gauge transformations (\ref{GaugeTrans}), and six from the invariance of the Lagrangian $\L$ under local Lorentz transformations. The twenty four original components of the Fierz tensor are then reduced to only two, as appropriate for a massless spin-2 field.

\subsection{Relation to the metric approach}

Let us consider the inertial frame $e'_a$ endowed with a cartesian coordinate system. In this case all connections vanish, and according to Eq.~(\ref{todete}) the tetrad $e'^a{}_\mu$ satisfies the condition $\partial_\rho e'^a{}_\mu = 0$. Using this tetrad, we can define $\phi'^\rho{}_\mu := e'_a{}^\rho \, \phi'^a{}_\mu$. Since $\phi'^\rho{}_\mu$ is not in principle symmetric, the perturbation of the metric---which is usually supposed to represent a fundamental spin-2 field---is to be identified with the symmetric part of $\phi'^\rho{}_\mu$:
\be
\psi^\rho{}_\mu = \phi'^\rho{}_\mu + \phi'_\mu{}^\rho.
\ee
It is then easy to see that, in terms of $\psi^\rho{}_\mu$, the gauge transformation (\ref{GaugeTrans}) acquires the form,
\be
\psi^\rho{}_\mu \to \psi^\rho{}_\mu - \partial^\rho \xi_\mu(x)
- \partial_\mu \xi^\rho(x),
\ee
where $\xi^\mu(x) = \xi^a(x) \, e'_a{}^\mu$. The Bianchi identity (\ref{BI}), on the other hand, is seen to be trivially satisfied, whereas the field equation (\ref{eom linear}) assumes the form
\begin{equation}
\Box\left(\delta^\mu{}_\lambda \, \psi - \psi^\mu{}_\lambda
\right) - \partial_\lambda \partial^\mu \, \psi - \delta^\mu{}_\lambda \, \partial_\nu
\partial_\rho \, \psi^{\nu \rho} + \partial_\nu \partial^\mu \,\psi^\nu{}_\lambda + 
\partial_\lambda \partial_\rho \, \psi^{\rho \mu} = 0,
\end{equation}
with $\psi = \psi^\alpha{}_\alpha$. This is precisely the linearized Einstein equation \cite{deser1}, which means that in absence of gravity the teleparallel-based approach is totally equivalent to the usual general relativity-based approach to the spin-2 field. This means that $\phi^a{}_\rho$ and $\psi_{\mu \nu}$ are the same physical field, which in the massless case is well known to represent waves with helicity 2. The teleparallel approach, however, is much more elegant and simple in the sense that it is similar to the spin-1 electromagnetic theory---an heritage of the (abelian) gauge structure of teleparallel gravity. In addition, it allows for a precise distinction between gauge transformations---local translations in the tangent space---and spacetime coordinate transformations. Furthermore, in contrast to the usual metric approach, the field $\phi^a{}_\rho$, as well as its gauge transformations, are not restricted to be infinitesimal.

\section{Spin-2 field in the presence of gravitation}
\label{s2plusGra}

\subsection{Gravitational coupling prescription}
\label{CouPres}

In absence of gravitation, as we have seen in the previous section, the algebraic index of $\phi^a{}_\rho$ can be transformed into a spacetime index through contraction with the tetrad field, and vice-versa. In the presence of gravitation, this index transformation can lead to problems with the coupling prescription. To begin with we note that, because $\phi^a{}_\rho$ is a vector field assuming values in the Lie algebra of the translation group, $\phi_\rho = \phi^a{}_\rho P_a$, the algebraic index ``$a$'' is not an ordinary vector index. It is actually a gauge index which, due to the ``external'' character of translations, happens to be similar to the usual, true vector index ``$\rho$''. To understand better this difference, let us consider a scalar field $\phi$. As is well known, its gravitational coupling prescription is trivial:
\be
\partial_\mu \phi \to \partial_\mu \phi.
\ee
Its interaction with gravitation comes solely from the tetrad replacement
\be
e^a{}_{\mu} \to h^a{}_{\mu} = e^a{}_{\mu} + B^a{}_\mu,
\label{GraviTetra}
\ee
or equivalently, from the metric replacement
\be
\eta_{\mu \nu} = e^a{}_\mu e^b{}_\nu \, \eta_{ab} \to
g_{\mu \nu} = h^a{}_\mu h^b{}_\nu \, \eta_{ab}.
\ee
The crucial point is to note that this is true independently of whether the scalar field is or is not a translational-valued field $\phi = \phi^a P_a$. We see in this way that translational gauge indices are irrelevant for the gravitational coupling prescription.

Based on the above considerations, in the class of frames $h'_a$ in which the inertial connection $\Aw^a{}_{b \mu}$ vanishes, the gravitational coupling prescription of the spin-2 field $\phi_\rho = \phi^a{}_\rho P_a$ is written in the form
\be
\partial_\mu \phi'_\rho \to
\partial_\mu \phi'_\rho - \left( \Gammaw^\lambda{}_{\rho\mu} -
\Kw^\lambda{}_{\rho\mu}\right) \phi'_\lambda.
\label{TCPgf0}
\ee
In components, it reads
\be
\partial_\mu \phi'^a{}_\rho \to
\partial_\mu \phi'^a{}_\rho - \left( \Gammaw^\lambda{}_{\rho\mu} -
\Kw^\lambda{}_{\rho\mu}\right) \phi'^a{}_\lambda.
\label{TCPgf1}
\ee
{Of course, because of the identity
\be
\Gammaw^\lambda{}_{\rho \mu} - \Kw^\lambda{}_{\rho \mu} = \Gammabol^\lambda{}_{\rho \mu},
\label{relation0}
\ee
with $\Gammabol^\lambda{}_{\rho \mu}$ the Levi-Civita connection of the metric $g_{\mu \nu}$, this coupling prescription coincides with the coupling prescription of general relativity. This is a key point of the equivalence between general relativity and teleparallel gravity \cite{TSC}.}

Let us consider now the coupling prescription in an arbitrary frame. Under a local Lorentz transformation, the ordinary derivative of $\phi'^a{}_\rho$ transforms according to
\be
\partial_\mu \phi'^a{}_\rho \to \Dw_\mu \phi^a{}_\rho =
\partial_\mu \phi^a{}_\rho + \Aw^a{}_{b \mu} \, \phi^b{}_\rho.
\ee
In a general Lorentz frame, therefore, the gravitational coupling prescription of a fundamental spin-2  field $\phi^a{}_\rho$ is written as
\be
\partial_\mu \phi^a{}_\rho \to 
\partial_\mu \phi^a{}_\rho + \Aw^a{}_{b \mu} \, \phi^b{}_\rho - \left( \Gammaw^\lambda{}_{\rho\mu} - \Kw^\lambda{}_{\rho\mu}\right) \phi^a{}_\lambda.
\label{TCPgf}
\ee
This coupling prescription provides different connection-terms for each index of $\phi^a{}_\rho$: whereas the algebraic index is connected to inertial effects only, the spacetime index is connected to the gravitational coupling prescription. It constitutes one of the main differences of the teleparallel-based approach in relation to the usual metric approach based on general relativity. In fact, the latter considers both indices of the spin-2 variable $\psi_{\mu \nu}$ on an equal footing, leading to a coupling prescription that breaks the gauge invariance of the spin-2 theory. It is important to note that, as a simple inspection shows, the coupling prescription (\ref{TCPgf}) cannot be rewritten in terms of the connections $\Abol^a{}_{b \mu} $ and $\Gammabol^\lambda{}_{\rho\mu}$ of general relativity. 

\subsection{Field strength and Bianchi identity}

Let us now apply the gravitational coupling prescription (\ref{TCPgf}) to the free theory. To begin with we notice that, because the gauge parameter $\xi^a$ has an algebraic index only, the gauge transformation (\ref{GenGT}) does not change in the presence  of gravitation:
\be
\phi^a{}_\mu \to \phi^a{}_\mu - \Dw_\mu \xi^a(x).
\label{GenGTbis}
\ee
On the other hand, considering that the connection (\ref{relation0}) is symmetric in the last two indices, we see that the field strength $F^a{}_{\mu \nu}$ does not change in the presence of gravitation: 
\be
F^a{}_{\mu \nu} = \Dw_\mu \phi^a{}_{\nu} - \Dw_\nu \phi^a{}_{\mu}.
\label{FS0}
\ee
Due to the fact that the teleparallel Fock-Ivanenko derivative $\Dw_\mu$ is commutative, the Bianchi identity also remains unchanged,
\be
\Dw_\rho F^a{}_{\mu \nu} + \Dw_\nu F^a{}_{\rho \mu} + \Dw_\mu F^a{}_{\nu \rho} = 0.
\label{BI3bisbis}
\ee
Since $\varepsilon^{\lambda \rho \mu \nu}$ is a density of weight $\omega = - 1$, in the presence of gravitation the Levi-Civita {\em tensor} is given by $h \varepsilon^{\lambda \rho \mu \nu}$, and the Bianchi identity can be rewritten in the form
\be
\Dw_\rho (h \, \varepsilon^{\lambda \rho \mu \nu} \, F^a{}_{\mu \nu}) = 0.
\ee
This is similar to what happens to the electromagnetic field in the presence of gravitation.

\subsection{Lagrangian and field equation}

Analogously to the flat background case, the Lagrangian of the spin-2 field in the presence of gravitation can be obtained from the teleparallel Lagrangian (\ref{tela}) by replacing the teleparallel torsion $\Tw^a{}_{\mu \nu}$ by the spin-2 field strength $\sqrt{k} \, F^a{}_{\mu \nu}$. The result is
\be
{\mathcal L} =
\frac{h}{4} \, {\mathcal F}_a{}^{\mu\nu} \, F^a{}_{\mu\nu},
\label{TeleLa}
\ee
where 
\be
{\mathcal F}_a{}^{\mu \nu} = h_a{}^\rho \, {\mathcal K}^{\mu \nu}{}_\rho - h_a{}^\mu \, h_b{}^\sigma F^{b \nu}{}_\sigma + h_a{}^\nu \, h_b{}^\sigma F^{b \mu}{}_\sigma
\ee
is the gravitationally-coupled Fierz tensor, with
\be
{\mathcal K}^{\mu \nu}{}_\rho = \onehalf \left(h_a{}^\nu \, F^{a \mu}{}_\rho +
h^a{}_\rho \, F_a{}^{\mu \nu} - h_a{}^\mu \, F^{a \nu}{}_\rho \right)
\ee
the corresponding spin-2 analogous of the coupled contortion. We notice in passing that this Lagrangian is invariant under the gauge transformation (\ref{GenGT}). It is furthermore invariant under local Lorentz transformation $h'^a{}_\mu = \Lambda^a{}_b(x) \, h^a{}_\mu$ of the frames.

Performing variations in relation to $\phi^a{}_\rho$, we get
\be
\Dw_\mu {\mathcal F}_a{}^{\rho \mu} +
( \Gammaw^\mu{}_{\nu\mu} - \Kw^\mu{}_{\nu\mu}) \, {\mathcal F}_a{}^{\rho \nu} = 0.
\ee
Using the identity
\be
\partial_\mu h = h \, \Gammabol^\mu{}_{\lambda \mu} \equiv
h \left(\Gammaw^\mu{}_{\lambda \mu} - \Kw^\mu{}_{\lambda \mu} \right)
,
\label{delh}
\ee
it can be rewritten in the form
\be
\Dw_\mu (h {\mathcal F}_a{}^{\rho \mu}) = 0.
\label{FE1}
\ee
This is the field equation of a fundamental spin-2 field in the presence of gravitation, as seen from the general frame $h^a{}_\mu$. It can also be obtained from the free field equation (\ref{eom linear2}) by applying the gravitational coupling prescription (\ref{TCPgf}). It is important to remark that, on account of the commutativity of the covariant derivative $\Dw_\mu$, the gravitationally-coupled theory is gauge invariant and, like the free theory, has the correct number of independent components.

\section{Spin-2 field as source of gravitation}
\label{s2Source}

Let us consider now the total Lagrangian
\be
{\mathcal L}_t = \Lw + {\mathcal L},
\ee
where $\Lw$ is the teleparallel Lagrangian (\ref{tela}), and ${\mathcal L}$ is the Lagrangian (\ref{TeleLa}) of a spin-2 field in the presence of gravitation. The corresponding field equation is
\be
\partial_\sigma (h \sw_a{}^{\rho \sigma}) -
k \, h \, (\tw_{a}{}^{\rho} + \iw_a{}^{\rho}) = k \, h \, \theta_{a}{}^{\rho},
\label{fe22}
\ee
where $\tw_{a}{}^{\rho}$ is the gravitational energy-momentum tensor, $\iw_a{}^{\rho}$ is the energy-momentum pseudotensor of inertia, and
\be
\theta_a{}^\rho \equiv - \frac{1}{h} \frac{\delta {\mathcal L}}{\delta h^a{}_\rho} =
h_a{}^\nu \, {\mathcal F}_c{}^{\mu \rho} \, F^c{}_{\mu \nu} - \frac{h_a{}^\rho}{h} \, {\mathcal L}
\ee
is the spin-2 field source energy-momentum tensor. Observe that
\be
\theta_\rho{}^\rho \equiv h^a{}_\rho \, \theta_a{}^\rho = 0,
\ee
as it should be for a massless field. Furthermore, from the invariance of ${\mathcal L}$ under general co\-ordinate transformation, it is found to satisfy the usual (general relativity) covariant conservation law \cite{weinberg}
\be
\Dbol_\rho (h \theta_a{}^\rho) \equiv \partial_\rho (h \theta_a{}^\rho) -
\Abol^b{}_{a \rho} \,  (h \theta_b{}^\rho) = 0.
\ee
Due to the anti-symmetry of the superpotential in the last two indices, we see from the field equation (\ref{fe22}) that the total energy-momentum density is conserved in the ordinary sense:
\be
\partial_\rho [h (\iw_a{}^\rho + \tw_a{}^\rho + \theta_{a}{}^{\rho})] = 0.
\ee

The field equation (\ref{fe22}) can be rewritten in the form
\be
\partial_\sigma (h \sw_a{}^{\rho \sigma}) -
k \, h \, \iw_a{}^{\rho} = k \, h \, (\tw_a{}^\rho + \theta_{a}{}^{\rho}),
\label{fe53}
\ee
where the right-hand side represents the true gravitational field source. We recall that the energy-mo\-men\-tum density of inertia, although entering the total energy-momentum conservation, is not source of gravitation, and accordingly must remain in the left-hand side of the field equation. 
Now, as already discussed in section~\ref{TG}, the two terms on the left-hand side form a covariant derivative. This allows the field equation (\ref{fe53}) to be rewritten in the form
\be
\Dw_\sigma (h \sw_a{}^{\rho \sigma}) = k \, h \, (\tw_a{}^\rho + \theta_{a}{}^{\rho}).
\label{fe21}
\ee
Considering that the covariant derivative $\Dw_\rho$ is commutative, the true source of gravitation is found to be conserved in the covariant sense:
\be
\Dw_\rho [h (\tw_a{}^\rho + \theta_{a}{}^{\rho})] = 0.
\label{CoCoLaw}
\ee
This property ensures the consistency of the theory in the sense that no constraints on the background spacetime geometry show up.

\section{Final remarks}
\label{FR}

Due to the fact that it describes the gravitational interaction through a geometrization of spacetime, general relativity is not, strictly speaking, a field theory in the usual sense of classical fields. On the other hand, owing to its gauge structure, teleparallel gravity does not geometrize the gravitational interaction, and for this reason it is much more akin to a field theory than general relativity. When looking for a field theory for the spin-2 field, therefore, it seems far more reasonable to use teleparallel gravity as paradigm. Accordingly, instead of a symmetric second-rank tensor $\psi_{\mu \nu}$, the spin-2 field is assumed to be represented by a spacetime (world) vector field assuming values in the Lie algebra of the translation group. Its components $\phi^a{}_\mu$, like the gauge potential of teleparallel gravity (or the tetrad), represent a set of four spacetime vector fields.

In absence of gravitation, the resulting spin-2 field theory naturally emerges in the Fierz formalism, and turns out to be structurally similar to electromagnetism, a gauge theory for the $U(1)$ group. In fact, in addition to satisfy a dynamic field equation, the spin-2 field is found to satisfy also a Bianchi identity, which is related to the dynamic field equation by duality transformation. Furthermore, the gauge and the local Lorentz invariance of the theory provide the correct number of independent components for a massless spin-2 field. Upon contraction with the tetrad, the translational-valued field can be transformed into a symmetric second-rank tensor field, in which case the theory reduces to the usual metric-based spin-2 theory already discussed in the literature \cite{deser2}. This shows that $\phi^a{}_\mu$ and $\psi_{\mu \nu}$ represent the same physical field, and that both approaches are equivalent in absence of gravitation. The teleparallel-based construction, however, can be considered to be more elegant in the sense that it has a gauge structure, it presents duality symmetry, and it allows for a precise distinction between gauge transformations---local translations in the tangent space---and spacetime coordinate transformations. Furthermore, similar to electromagnetism, neither $\phi^a{}_\rho$ nor its gauge transformations are restricted to be infinitesimal.

In the presence of gravitation, the teleparallel-based approach differs substantially from the usual metric approach. The reason is that the index $``a$'' of the translational-valued field $\phi^a{}_\rho$ is not an ordinary vector index, but a gauge index. As such, it is irrelevant for the gravitational coupling prescription, as discussed in section~\ref{CouPres}. This point is usually overlooked in the metric approach, which considers both indices of the spin-2 field $\psi_{\mu \nu}$ on an equal footing. As a result, the ensuing gravitational coupling prescription is found to break the gauge invariance of the theory. When the correct coupling prescription is used, a sound gravitationally-coupled spin-2 field theory emerges, which is quite similar to the gravitationally-coupled electromagnetic theory. Furthermore, it is both gauge and local Lorentz invariance, and it preserves the duality symmetry of the free theory.

As we have seen, the gravitational field equation of teleparallel gravity can be written in the form
\be
\Dw_\sigma (h \sw_a{}^{\rho \sigma}) = 
k \, h \, (\tw_a{}^\rho + \theta_{a}{}^{\rho}),
\label{fe23}
\ee
where the right-hand side represents the true source of gravitation---the sum of the gravitational and the source energy-momentum {\em tensors}. The inertial energy-momentum density, which is not source of gravitation, is implicit in the covariant derivative of the left-hand side---as can be seen from Eq.~(\ref{DotFI}). A crucial point of this equation is that, owing to the fact that the teleparallel spin connection $\Aw^a{}_{b \mu}$ is purely inertial, the covariant derivative $\Dw_\sigma$ is commutative. Taking into account the anti-symmetry of the superpotential in the spacetime indices, we obtain the divergence identity
\be
\Dw_\rho \Dw_\sigma (h \sw_a{}^{\rho \sigma}) = 0,
\ee
which is consistent with the covariant conservation law (\ref{CoCoLaw}). This property, together with the gauge and local Lorentz invariance, render the gravitationally-coupled spin-2 theory fully consistent. In the context of general relativity, whose spin connection represents both inertia and gravitation, the inertial part of the gravitational energy-momentum pseudotensor cannot be separated, and consequently the gravitational field equation cannot be written in a form equivalent to (\ref{fe23}). In the context of general relativity, therefore, no consistent gravitationally-coupled spin-2 field theory can be obtained.

\section*{Acknowledgments}
The authors would like to thank Yu.\ Obukhov for useful discussions. They would like to thank also FAPESP, CNPq and CAPES for partial financial support.

\end{document}